\documentclass[twocolumn]{aastex631}
\usepackage{enumitem}

\setlist[enumerate]{itemsep=0pt,parsep=0pt}
\newcommand{\dsct}{$\delta$ Sct }
\newcommand{\gdor}{$\gamma$ Dor }
\newcommand{\msun}{$M_{\odot}$ }
\newcommand{\dshybrid}{$\delta$ Sct$| \gamma$ Dor }
\newcommand{\gdhybrid}{$\gamma$ Dor$| \delta$ Sct }
\newcommand{\kapa}{$\kappa$ }

\begin{document}

\title{The rotation properties of \dsct  and \gdor stars }
%\footnote{Released on March, 1st, 2024}} 

\author{Jiyu Wang}
\affiliation{CAS Key Laboratory of Optical Astronomy, National Astronomical Observatories, Chinese Academy of Sciences, Beijing 100101, China}
\affiliation{School of Astronomy and Space Science, University of the Chinese Academy of Sciences, Beijing, 100049, China}

\author{Xiaodian Chen}
\affiliation{CAS Key Laboratory of Optical Astronomy, National Astronomical Observatories, Chinese Academy of Sciences, Beijing 100101, China}
\affiliation{School of Astronomy and Space Science, University of the Chinese Academy of Sciences, Beijing, 100049, China}
\affiliation{Institute for Frontiers in Astronomy and Astrophysics, Beijing Normal University, Beijing 102206, China}
\affiliation{Department of Astronomy, China West Normal University, Nanchong, 637009, China}

\author{Licai Deng}
\affiliation{CAS Key Laboratory of Optical Astronomy, National Astronomical Observatories, Chinese Academy of Sciences, Beijing 100101, China}
\affiliation{School of Astronomy and Space Science, University of the Chinese Academy of Sciences, Beijing, 100049, China}
\affiliation{Department of Astronomy, China West Normal University, Nanchong, 637009, China}

\author{Jianxing Zhang}
\affiliation{CAS Key Laboratory of Optical Astronomy, National Astronomical Observatories, Chinese Academy of Sciences, Beijing 100101, China}
\affiliation{School of Astronomy and Space Science, University of the Chinese Academy of Sciences, Beijing, 100049, China}

\author{Weijia Sun}
\affiliation{Leibniz-Institut für Astrophysik Potsdam (AIP), An der Sternwarte 16, 14482 Potsdam, Germany}

\correspondingauthor{Jiyu Wang, Xiaodian Chen}
\email{wangjy@bao.ac.cn, chenxiaodian@nao.cas.cn}

\begin{abstract}
Based on the LAMOST spectroscopy and TESS time-series photometry, we have obtained a main-sequence star sample of $\delta$ Scuti and $\gamma$ Doradus stars. The sample includes 1534 \dsct stars, 367 \gdor stars, 1703 \dshybrid stars, 270 \gdhybrid stars, along with 105 `dsct candidates' and 32 `gdor candidates'. After correcting for projection effects, we derived the equatorial rotational velocity distribution for \dsct  and \gdor stars and compared it with that of normal stars. The rotational velocity distributions of \dsct and \gdor stars are extremely similar, with the only difference potentially due to  the rotational variable stars that have not been completely removed.  In contrast, the rotational velocity distribution of normal stars is more dispersed compared to pulsating stars. Additionally, the peak rotational velocity of the pulsating stars is about 10 km s$^{-1}$ higher than that of normal stars. Unlike the normal stars, which show a monotonic increase in peak velocity with mass between 1.8 and 2.5 \msun, the rotational velocity distribution of \dsct stars does not exhibit a strong mass dependence.  We also found that normal stars accelerate during the late main-sequence evolutionary phase, while \dsct stars decelerate. Furthermore, there may still be unclassified stars with diverse rotational properties in the normal star sample compared to the \dsct stars, which is likely to be an important contributor to the broader dispersion observed in its rotational velocity distribution. The photometric amplitude in \dsct stars is modulated with rotational velocity, with high-amplitude stars typically rotating slowly and low-amplitude stars showing a broad distribution of rotational velocities. 

%This may be attributed to the significant influence of rotation on radial pulsations, where only slowly rotating \dsct stars can generate high-amplitude radial pulsations, while non-radial pulsations tend to inhibit the generation of high-amplitude radial pulsations.

\end{abstract}


\section{Introduction}\label{sec:intro}
The $\delta$ Scuti and $\gamma$ Doradus pulsators, are located at the intersection of the classical instability strip and the zero-age main sequence. \dsct stars falls into the category of short-period variable stars with a mass range of 1.5 to 2.5 \msun  (0.01 d $ < P < $ 0.25 d), showing typical V-band amplitudes ranging from 3 mmag to 0.9 mag  \citep[GCVS;][]{2017ARep...61...80S}. The majority of \dsct stars exhibit multi-period pulsations characterized by high-amplitude radial pulsations and low-amplitude non-radial pulsations. These pulsations are mainly induced by the opacity mechanism (the \kapa mechanism) in the He II ionization zone \citep{2010aste.book.....A}, usually resulting in low-radial order, low-degree modes. In contrast, \gdor stars is categorized as a type of long-period variable star with a mass range of 1.2 to 2.0 \msun (0.3 d $ <P< $ 3 d), typically showing ${\rm Amp}_{V}$ below 0.1 mag. The pulsations observed in \gdor stars are primarily high-order non-radial g-mode pulsations excited by the convective flux blocking mechanism \citep{2005A&A...435..927D,2010ApJ...713L.192G}. However, there exists an instability overlap region in the HR diagram for \dsct and \gdor stars, which are referred to as ‘hybrid stars’  \citep{2002MNRAS.333..262H,2005AJ....129.2026H,2011A&A...534A.125U,2018A&A...610A..17L,2022MNRAS.510.1748N}. In these stars, we observe pressure mode and gravity mode simultaneously.

 The evolution of the stellar rotation profile is driven by the long-term evolution of angular momentum loss and redistribution, which begins during the stellar formation phase \citep{1997ApJ...478..569M}, influencing various aspects of stellar evolution \citep{2000ARA&A..38..143M}. The velocity distribution of stellar rotation plays a crucial role in comprehending the evolution of stellar populations  \citep{1990ApJS...74..501P,2000ARA&A..38..143M}. At approximately 1.2 \msun, there is a significant contrast in the rotational speeds of stars  \citep{1970saac.book..385K}. More massive stars, which lack deep convective zones and strong magnetic fields that hinder their rotation, can reach velocities of several hundred kilometers per second \citep{1999A&A...349..189S,2021ApJ...921..145S}.

Stellar pulsations play a critical role in the transport of angular momentum within a star's interior, significantly impacting its surface rotation. In slowly pulsating B-type stars (SPB), g-mode pulsations driven by heat remove angular momentum from the surface and redistribute it to particular regions within the interior \citep{2018MNRAS.475..879T}. Similarly, in the convective cores of Be stars, g-mode and r-mode pulsations that are stochastically excited help in transferring angular momentum from the core to the surface \citep{2020A&A...644A...9N}.

Notably, \dsct and \gdor stars, which have a specific mass range, typically exhibit strong rotation. Rotation leads to an increase in the degeneracy of non-radial pulsations, which in turn causes  frequency splitting. \citep{2010aste.book.....A}. The degree of  frequency splitting in the periodogram of non-radial pulsation peaks is directly linked to the rotational speed \citep{2010aste.book.....A,2017ApJ...838...31C}. Observations suggest a correlation between the rotation and the photometric amplitudes of pulsating stars \citep{2014ApJ...781...88A,2023A&A...674A..36G}. In the case of \dsct stars, the projected rotational velocities of high-amplitude \dsct stars generally do not exceed 30 km s$^{-1}$, while normal \dsct stars display a wider range of rotational velocities \citep{2007CoAst.150...25B,2013AJ....145..132C}.  Undoubtedly, there exists a intricate coupling between stellar rotation and pulsation, exerting profound effects on stellar evolution. However, our current understanding of this phenomenon remains limited. Therefore, we employ variables such as \dsct and \gdor stars, which exhibit similar characteristics but differ in the pulsation modes, to investigate the  influence between rotation and pulsations.

To determine the pulsation frequencies of variable stars, it is essential to have uninterrupted data sets characterized by a high signal-to-noise ratio and precise frequency resolution \citep{2021RvMP...93a5001A,2022ARA&A..60...31K}. In recent decades, a plethora of space telescopes such as WIRE \citep{2005ApJ...619.1072B}, MOST \citep{2007CoAst.150..333M}, CoRoT \citep{2009A&A...506..411A},  Kepler \citep{2010Sci...327..977B,2010ApJ...713L..79K,2010ApJ...713L.192G}, TESS \citep{2015JATIS...1a4003R}, among others, have supplied copious amounts of  precise and detailed data through continuous long-term photometric observations, enabling the detection of pulsation modes in variable stars. Furthermore, ASAS-SN \citep{2020MNRAS.493.4186J}, ATLAS \citep{2018AJ....156..241H}, WISE \citep{2018ApJS..237...28C}, and ZTF \citep{2020ApJS..249...18C} have also contributed significant datasets for comparative reference purposes.

The structure of the paper is as follows. In Section \ref{sec:data}, we discuss the data sample and the treatment of the contaminants. We classify the resulting catalog in Section \ref{sec:classify}. Section \ref{sec:rotation} describes how the equatorial rotation velocity distribution of the sample stars is obtained, and Section \ref{sec:discussion} discusses its influencing factors. Section \ref{sec:amp} addresses the relationship between amplitude and rotation in sample stars. Finally, Section \ref{sec:conclusion} provides a summary.

\section{Data}\label{sec:data}
 A catalog of 40034 early-type main-sequence stars, with parameters such as $\mathrm{T_{eff}}$, $\log g$, $v\sin i$, [M/H], was compiled from the medium-resolution survey  ($R \sim 7500$) conducted by LAMOST \citep{2021ApJS..257...22S}. By restricting the mass to below 2.5 \msun and matching with the TESS input catalog v8.2 \citep{unknown}, we compiled a catalog containing 18,968 stars with TIC identifiers.

In order to compile a pure “normal” (single, non-chemically peculiar, and non-cluster member) variable star catalog, it is imperative to eliminate contaminants.

Binaries, particularly close binary systems, can significantly impact the rotational evolution of stars due to tidal forces, mass transfer, and stellar mergers between companion stars \citep{2013ApJ...764..166D}. Tidal forces facilitate the transfer of stellar spin angular momentum to the orbital angular momentum of the binary system, gradually achieving orbital synchronization. Mass transfer involves the exchange of angular momentum between companion stars and similarly influences stellar rotation. 

Chemically peculiar (CP) stars can be categorized into four groups: non-magnetic metallic line (Am) stars, magnetically peculiar (mAp) stars, stars with enhanced Hg II and Mn II (HgMn), and He-weak stars \citep{1974ARA&A..12..257P}. Various types of CP stars exhibit various rotational distributions, but most CP stars exhibit slow rotation. 

According to research conducted by \cite{2006ApJ...648..580H}, it was observed that the rotational velocity distribution of stars within clusters deviates from that of field stars, manifesting a reduced prevalence of slowly rotating B-type stars among cluster members compared to those in the field. Furthermore, the study by \cite{2018A&A...612L...2K} revealed that the distribution of rotation axis inclinations among cluster member stars shows a distinct preference for a non-isotropic distribution, diverging from the isotropic distribution typically observed in field stars. Therefore, we exclude cluster member stars to obtain a clean sample of field stars.

Some contaminants within the sample have been classified and annotated by \citet{2021ApJS..257...22S}.  Additionally, after cross-matching with SIMBAD and excluding contaminant stars identified as spectroscopic binary (SB*), chemically peculiar star (Pec*), RS Canum Venaticorum variable (RSCVn), rotating variable (RotV*), $\alpha^2$ Canum Venaticorum variable (RotV*alf2CV), and eclipsing binary (EB), a total of 17,955 stars remained. Subsequently, we cross-matched with the Gaia DR3 variable star catalog \citep{2023A&A...674A..13E,2023A&A...674A..14R} and identified 2,140 stars. By excluding stars classified as eclipsing binary (ECL), $\alpha^2$ Canum Venaticorum (ACV), (magnetic) chemically peculiar (MCP, CP), rapidly oscillating Am/Ap-type (ROAM, ROAP), SX Arietis (SXARI) star (ACV$|$CP$|$MCP$|$ROAM$|$ROAP$|$SXARI), and RS Canum Venaticorum variable (RS), we are left with 17,779 stars. Finally, we consider the contaminated stars marked by \citet{2021ApJS..257...22S}. After removing all known contaminants, we obtain a pure star catalog of 15421 stars.

We employ the LightKurve package to retrieve the latest time-series data from TESS \citep{2018ascl.soft12013L} and identify 13601 sample stars. Next, we use the Lomb-Scargle method to determine the dominant period \citep{1976Ap&SS..39..447L,1982ApJ...263..835S}. We discard all data with a signal-to-noise ratio below 10, leaving 7800 sample stars. We then performed a fourth-order Fourier fit on the light curves of these stars and fold them at twice the period. This procedure results in 4011 well-fitted stars, 2511 poorly fitted stars, and 1278 stars with asymmetric fits. Stars with asymmetric light curves are usually close binary stars, including EB and EW type eclipsing binaries, as well as ELL type rotating variable stars. Thus, after excluding all identified contaminants, we compile a catalog containing 4011 pulsation variable stars.

\begin{figure}[h]
    \centering
    \includegraphics[width=1\linewidth]{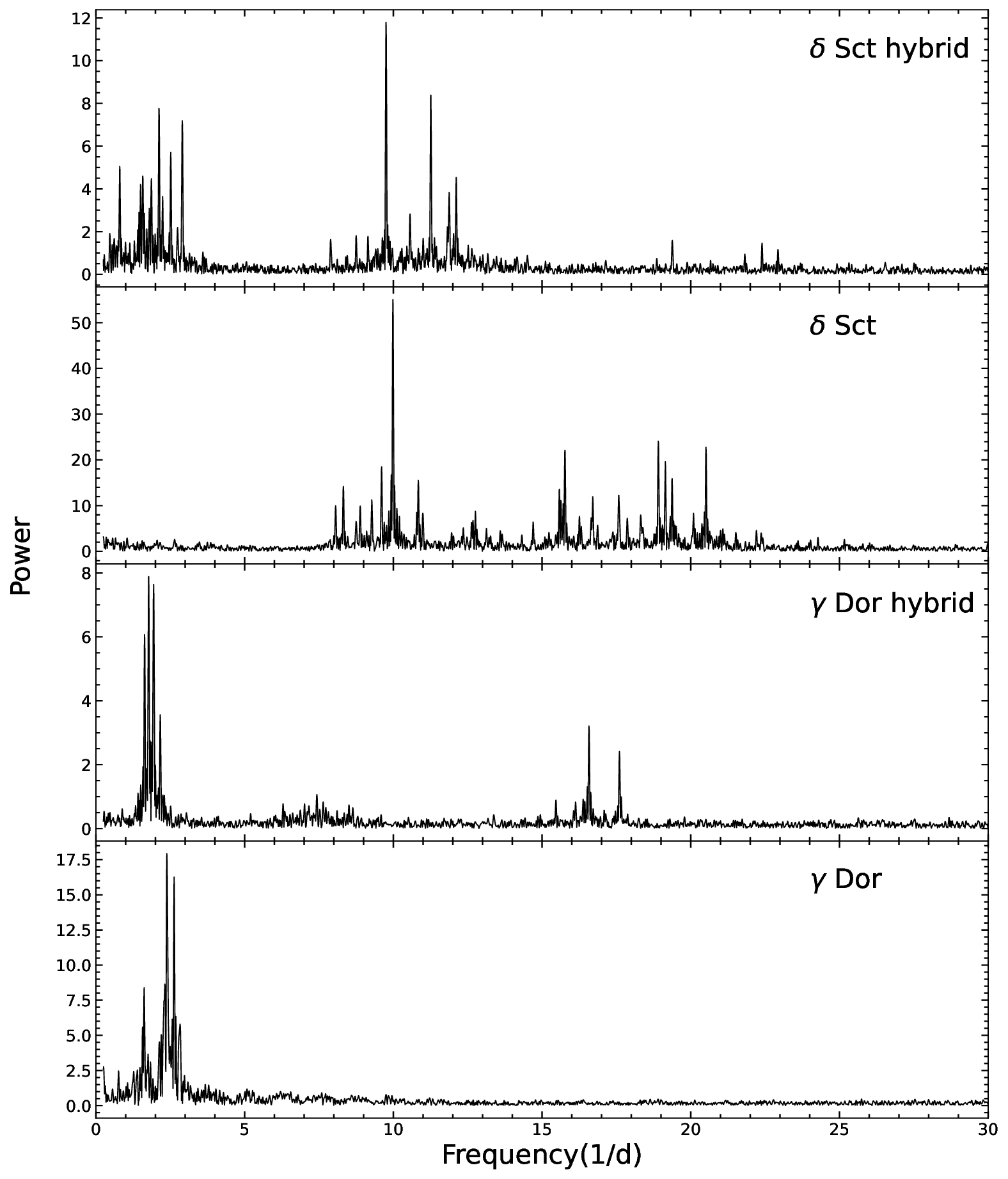}
    \caption{Frequency spectrum data derived from the TESS photometry. We classify hybrid stars based on the dominant pulsation modes into two categories, namely \dshybrid and \gdhybrid stars.}
    \label{fig:1}
\end{figure}

\begin{table*}[htbp]
\centering
\caption{\label{t1}Catalog of Rotational Properties for \dsct and \gdor Stars \footnote{This table is available in its entirety in machine-readable form.}}
\begin{tabular}{lcccccccccl}
\hline
\hline
TIC      &RA     &DEC   &log L  &$\mathrm{T_{eff}}$     &$v\sin i$   &period & Amp & ... &vari\_type   \\
&deg &deg &$\quad \log L_{\odot}$ &K &km s$^{-1}$ &days &mmag & & \\
\hline
1842330 & 78.85565 & 33.284946 & 1.07 ± 0.02 & 8166 ± 79   & 81 ± 3   & 1.33405 & 2.57  & ... & gdor           \\
2015042 & 79.23069 & 32.990090 & 0.99 ± 0.02 & 7357 ± 78   & 18 ± 3   & 0.06521 & 1.69  & ... & dsct           \\
2095379 & 79.27057 & 32.063739 & 1.63 ± 0.03 & 9727 ± 222  & 115 ± 9  & 0.12038 & 0.93  & ... & dsct           \\
2096047 & 79.27972 & 31.664515 & 1.02 ± 0.02 & 7081 ± 50   & 72 ± 2   & 0.08845 & 5.92  & ... & dsct           \\
2146051 & 79.51305 & 35.312433 & 1.37 ± 0.02 & 7750 ± 88   & 216 ± 4  & 0.10069 & 6.28  & ... & dsct hybrid    \\
2234995 & 79.67573 & 33.284152 & 1.60 ± 0.02 & 7404 ± 71   & 75 ± 3   & 0.08714 & 7.65  & ... & dsct hybrid    \\
2237237 & 79.66534 & 34.351272 & 1.76 ± 0.02 & 8553 ± 50   & 89 ± 2   & 1.26954 & 0.71  & ... & gdor           \\
2333031 & 79.76204 & 34.975885 & 1.49 ± 0.02 & 8253 ± 113  & 223 ± 5  & 1.82912 & 0.85  & ... & gdor           \\
2427014 & 79.88258 & 31.241129 & 1.10 ± 0.02 & 7321 ± 50   & 88 ± 2   & 0.04757 & 5.94  & ... & dsct           \\
2432832 & 79.92834 & 33.912190 & 1.32 ± 0.02 & 7472 ± 78   & 120 ± 3  & 0.05783 & 5.19  & ... & dsct           \\
2505113 & 80.07317 & 33.923281 & 0.81 ± 0.02 & 7280 ± 91   & 22 ± 4   & 0.41540 & 4.33  & ... & gdor hybrid    \\
2505443 & 80.02393 & 33.775741 & 1.90 ± 0.03 & 11783 ± 196 & 137 ± 8  & 2.16536 & 1.12  & ... & gdor           \\
2506993 & 80.09311 & 33.073012 & 1.45 ± 0.02 & 8530 ± 70   & 23 ± 3   & 0.55971 & 3.97  & ... & gdor           \\
2514753 & 80.21355 & 30.550876 & 1.14 ± 0.02 & 7739 ± 73   & 299 ± 2  & 0.06421 & 3.68  & ... & dsct hybrid    \\
2594144 & 80.13489 & 33.028923 & 1.38 ± 0.02 & 7122 ± 82   & 120 ± 4  & 0.08409 & 8.72  & ... & dsct           \\
2596395 & 80.13107 & 34.075335 & 0.89 ± 0.02 & 7232 ± 75   & 103 ± 3  & 0.04641 & 2.20  & ... & dsct hybrid    \\
2599498 & 80.17658 & 35.541374 & 1.43 ± 0.02 & 7973 ± 70   & 251 ± 3  & 0.04001 & 1.65  & ... & dsct hybrid    \\
2602315 & 80.30103 & 34.963391 & 1.19 ± 0.02 & 7297 ± 80   & 44 ± 4   & 0.04286 & 6.07  & ... & dsct           \\
2604466 & 80.31753 & 33.978959 & 1.18 ± 0.02 & 7857 ± 92   & 248 ± 4  & 0.26246 & 1.34  & ... & gdor hybrid    \\
2679334 & 80.38543 & 32.020190 & 1.68 ± 0.04 & 7420 ± 50   & 42 ± 2   & 0.09706 & 2.09  & ... & dsct hybrid    \\
2688166 & 80.43771 & 32.287047 & 1.45 ± 0.03 & 8317 ± 50   & 103 ± 2  & 0.05375 & 1.11  & ... & dsct           \\
2768898 & 80.46392 & 33.911677 & 1.35 ± 0.04 & 9010 ± 167  & 87 ± 7   & 1.61054 & 6.06  & ... & gdor           \\
2769016 & 80.42024 & 33.957888 & 0.97 ± 0.02 & 7482 ± 108  & 127 ± 5  & 0.05585 & 1.56  & ... & dsct           \\
2772225 & 80.43248 & 35.406328 & 1.24 ± 0.02 & 7640 ± 70   & 245 ± 3  & 0.28303 & 1.16  & ... & gdor hybrid    \\
2772723 & 80.51671 & 35.630302 & 1.35 ± 0.02 & 7721 ± 80   & 42 ± 4   & 0.05198 & 11.54 & ... & dsct hybrid    \\
2774095 & 80.60317 & 35.594136 & 1.80 ± 0.03 & 10529 ± 88  & 127 ± 4  & 0.30503 & 0.99  & ... & gdor hybrid    \\
2845289 & 80.68468 & 31.876645 & 1.09 ± 0.02 & 7326 ± 75   & 210 ± 3  & 0.05980 & 4.77  & ... & dsct hybrid    \\
2939323 & 81.98344 & 35.297221 & 1.37 ± 0.02 & 9035 ± 73   & 135 ± 3  & 0.04914 & 1.34  & ... & dsct           \\
2939656 & 81.99620 & 35.194288 & 1.45 ± 0.02 & 9005 ± 78   & 319 ± 2  & 0.03829 & 0.83  & ... & dsct           \\
2942914 & 82.03298 & 34.200718 & 1.51 ± 0.05 & 7588 ± 229  & 288 ± 10 & 0.50371 & 2.28  & ... & gdor           \\
3032338 & 82.16216 & 32.395807 & 1.26 ± 0.03 & 7643 ± 75   & 258 ± 3  & 0.08309 & 3.46  & ... & dsct hybrid    \\
3035199 & 82.11313 & 33.555553 & 1.32 ± 0.02 & 9338 ± 95   & 144 ± 4  & 0.51398 & 3.03  & ... & gdor           \\
3092576 & 82.20469 & 34.124446 & 1.71 ± 0.02 & 7256 ± 240  & 45 ± 10  & 0.22352 & 3.77  & ... & gdor hybrid    \\
3113047 & 82.33561 & 34.266714 & 1.52 ± 0.03 & 7518 ± 70   & 207 ± 3  & 0.20046 & 13.38 & ... & gdor hybrid    \\
3211665 & 82.35292 & 32.496196 & 0.99 ± 0.06 & 7222 ± 132  & 103 ± 6  & 0.15451 & 12.92 & ... & dsct hybrid    \\
3220915 & 82.41015 & 31.687580 & 1.60 ± 0.02 & 8688 ± 50   & 161 ± 2  & 0.10134 & 0.44  & ... & dsct candidate \\
3223698 & 82.47134 & 32.861108 & 0.93 ± 0.02 & 7759 ± 152  & 211 ± 7  & 0.04666 & 5.14  & ... & dsct hybrid    \\
3223876 & 82.41705 & 32.937868 & 1.38 ± 0.02 & 7011 ± 50   & 112 ± 2  & 0.12607 & 9.62  & ... & dsct           \\
3308518 & 82.47451 & 33.973989 & 1.32 ± 0.03 & 8396 ± 233  & 89 ± 10  & 0.04783 & 6.97  & ... & dsct candidate \\
3309447 & 82.41218 & 34.378615 & 1.04 ± 0.02 & 7865 ± 240  & 191 ± 10 & 0.02754 & 2.61  & ... & dsct           \\
3309468 & 82.50288 & 34.391836 & 1.07 ± 0.02 & 8075 ± 237  & 145 ± 10 & 0.02912 & 4.03  & ... & dsct           \\
3314193 & 82.51485 & 35.587909 & 1.33 ± 0.02 & 7875 ± 215  & 63 ± 9   & 0.05829 & 5.71  & ... & dsct           \\
3314388 & 82.58203 & 35.512653 & 1.09 ± 0.02 & 7484 ± 90   & 112 ± 4  & 0.04024 & 4.57  & ... & dsct hybrid    \\
3315879 & 82.60559 & 34.904769 & 1.10 ± 0.02 & 8540 ± 123  & 279 ± 5  & 0.05439 & 1.34  & ... & dsct hybrid    \\
3317072 & 82.61165 & 34.421500 & 1.21 ± 0.02 & 9104 ± 76   & 50 ± 3   & 0.43886 & 2.35  & ... & gdor hybrid    \\
3320341 & 82.62175 & 33.032390 & 1.35 ± 0.02 & 8419 ± 81   & 147 ± 4  & 0.06245 & 0.84  & ... & dsct           \\
3321435 & 82.57678 & 32.562486 & 1.17 ± 0.02 & 8471 ± 162  & 252 ± 7  & 0.03310 & 5.43  & ... & dsct hybrid    \\
3455644 & 82.69717 & 35.105331 & 1.20 ± 0.02 & 8786 ± 78   & 167 ± 3  & 0.03074 & 4.94  & ... & dsct           \\
3457292 & 82.69492 & 35.783197 & 1.46 ± 0.02 & 8515 ± 119  & 213 ± 5  & 0.07348 & 2.21  & ... & dsct hybrid    \\
...     & ...      & ...       & ...         & ...         & ...      & ...     & ...   & ... & ...     \\
\hline
\end{tabular}
\end{table*}

\section{classification}\label{sec:classify}
The \citet{2021ApJS..257...22S} sample of early-type main-sequence stars has a mass range of approximately 1.5-6 \msun. Once all contaminants are excluded, the sample is theoretically composed of only three types of pulsating stars: \dsct stars, \gdor stars, and SPB stars. SPB stars are a type of variable star with a mass distribution ranging from 3 to 8 \msun, exhibiting non-radial g-mode pulsations driven by the \kapa mechanism on the iron bump \citep{1991A&A...246..453W,2018A&A...612L...2K}. Above 2.5 \msun, early-type main sequence stars exhibit different rotation velocity distributions, showing a bimodal distribution \citep{2012A&A...537A.120Z,2021ApJ...921..145S}. Therefore, we constrain the mass of the sample to be below 2.5 \msun to ensure that only \dsct and \gdor stars remain in the sample.

\dsct and \gdor stars can be distinguished by their photometric periods, despite the overlap in their instability strips, and the existence of hybrid stars \citep{2005AJ....129.2026H}.  Typically, a division of 0.2 d is commonly used to separate \dsct and \gdor stars \citep{2011A&A...534A.125U}, though other limits like 0.25 d have also been applied \citep{2019MNRAS.490.4040A}. Nevertheless, it is important to recognize that due to stellar rotation, \gdor stars can show pulsation periods shorter than 0.2 d \citep{2013MNRAS.429.2500B}, and likewise, \dsct stars may exhibit periods longer than 0.2 d \citep{2010ApJ...713L.192G}. Therefore, depending solely on a 0.2 d benchmark to distinguish between \dsct and \gdor stars is insufficient for obtaining a clean and unambiguous sample \citep{2022A&A...666A.142S,2024A&A...688A..25S}.
We classify the sample stars into four groups based on the dominant period and the pulsation mode distribution.

 \begin{enumerate}
    \item \dsct stars: The dominant frequency is greater than 5 c/d, with at least two significant peaks ($S/N > 4$) above 5 c/d, and no more than one significant peak below 5 c/d.
    \item \dshybrid stars:  The dominant frequency is greater than 5 c/d, with at least two significant peaks both below and above 5 c/d
    \item \gdor stars: The dominant frequency is less than 5 c/d, with at least two significant peaks below 5 c/d, and no more than one significant peak above 5 c/d.  
    \item \gdhybrid stars: The dominant frequency is less than 5 c/d, with at least two significant peaks both below and above 5 c/d
\end{enumerate}

\begin{figure}[h]
    \centering
    \includegraphics[width=1\linewidth]{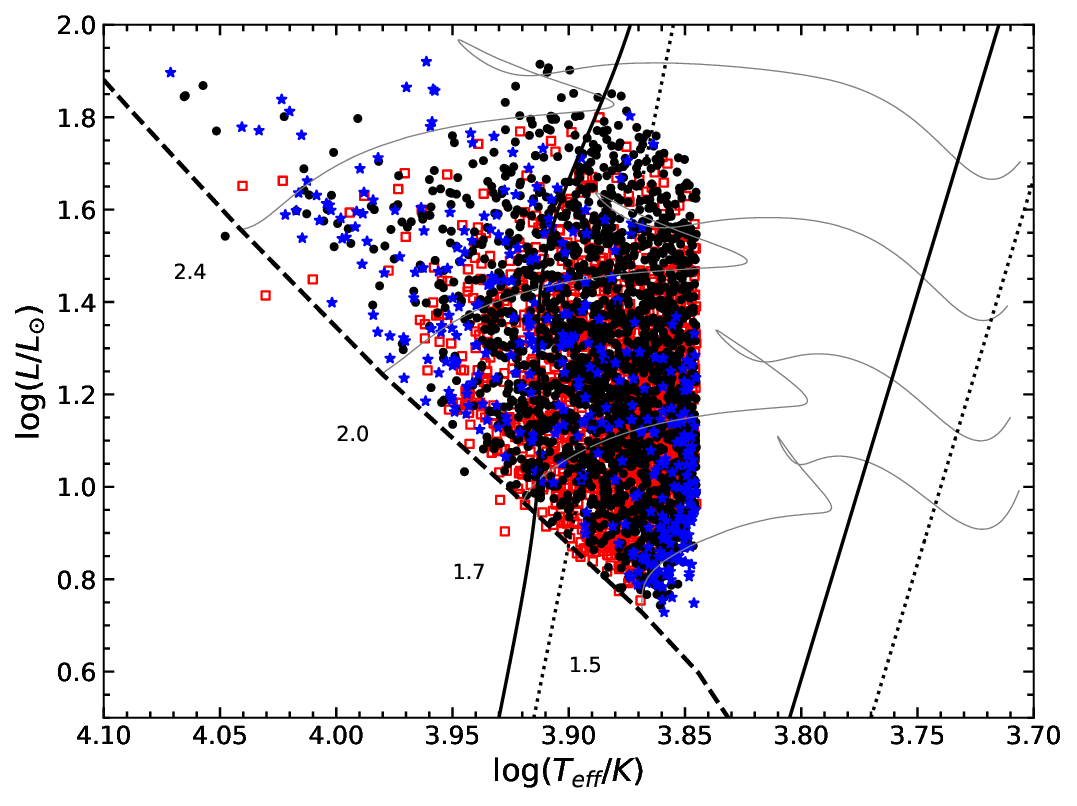}
    \caption{The figure shows the distribution of sample stars on the Hertzsprung-Russell diagram. The solid black line and dashed black line represent the theoretical instability strips boundaries for \dsct and \gdor stars, respectively, as derived from \citet{2016MNRAS.457.3163X}. The red open squares denote \dsct stars, blue asterisks denote \gdor stars, and black solid circles denote hybrid stars.}
    \label{fig:2}
\end{figure}

 Additionally, residual removal is necessary to reduce the influence of stronger peaks and ensure that the identified signal peaks are both independent and authentic. Given the occasional spurious strong peaks that occasionally appear at the low-frequency edge, close to the preset frequency threshold, we adopted an artificially experienced selection method to refine the data set, resulting in 130 stars previously classified as \dshybrid being reclassified to the \dsct category. In Fig.\hspace{0.25em}\ref{fig:1}, we display the frequency spectra for the four star categories, from which we can easily distinguish the g-mode and p-mode frequencies. According to our classification standards, we have identified 1,534 \dsct stars, 1,703 \dshybrid stars, 367 \gdor stars, and 270 \gdhybrid stars. Moreover, for sample stars that exhibit a single significant peak solely on the dominant frequency side, irrespective of the number of peaks on the opposite side, we tentatively categorize them into candidate classes. This includes 105 "dsct candidates" and 32 "gdor candidates". For the `DSCT candidates', these might indeed be genuien \dsct or \dshybrid stars, particularly as 23 of them have data from long-cadence (LC) photometry. For LC data, the Nyquist frequency is 24 c/d, implying that additional strong peaks may be present at higher frequencies. For ‘GDOR candidates’, we tend to consider them as stars with unresolved rotational modulation within the sample. We plot the distribution of sample stars on the Hertzsprung-Russell diagram, as well as the instability strips of \dsct and \gdor stars, as shown in Fig.\hspace{0.25em}\ref{fig:2}. The theoretical instability strips in the figure are from \citet{2016MNRAS.457.3163X}. Because the effective temperature limit of the stars in the sample are higher than 7000 K, they are mainly found clustered towards the bluer side close to the instability strip. It is evident that there are stars, regardless of whether they belong to the \dsct, \gdor, or hybrid categories, that are hotter than the blue edge of the instability strip. The range of \gdor and hybrid stars is more extensive compared to their instability strips  \citep{2019MNRAS.485.2380M}. This suggests that our knowledge about \gdor and hybrid stars is not yet comprehensive enough such that the theoretical instability strips fall short of providing sufficient restrictions.

A partial listing of the sample star catalog is provided in Table \ref{t1}, which includes parameters like $\mathrm{T_{eff}}$, $\log L$, and $v\sin i$ that are derived from the stellar catalog by \citep{2021ApJS..257...22S}. In this work, the primary period, the amplitude in the TESS band, and the classification of variable stars are determined. Additionally, the catalog includes three frequencies fitted during the classification process along with their signal-to-noise ratios, as well as classification data from Gaia DR3 and SIMBAD.

\section{rotation distribution}\label{sec:rotation}
In order to determine the equatorial rotation velocity of a star, it is necessary to correct the projected rotation velocity $v\sin i$ for inclination. However, except in specific scenarios such as binary stars with orbits synchronized by strong tidal forces or stars exhibiting rotation-modulated signals in their light curves, we cannot directly measure the inclination of a star's rotation. To address this, we assume that the distribution of rotational tilt angles of the star is isotropic, which provides a statistical correction for the effect of inclination angles \citep{1995ApJS...99..135A,2012A&A...537A.120Z}.

Under the assumption of random distribution along the rotational axis, we estimate the distribution of equatorial rotational velocity by the following method \citep{2007A&A...463..671R,2012A&A...537A.120Z}:
\begin{equation}
\Psi(\theta) = \int \gamma(v)P(\theta|v)dv = 
\int \gamma(v)\frac{\theta}{v}\frac{H(v-\theta)}{\sqrt{v^2-\theta^2}}dv
\end{equation}

In cases where $\theta = v\sin i$, $\psi(\theta)$ denotes the probability density function (PDF) achieved through Gaussian kernel smoothing, $\gamma(v)$ symbolizes the PDF of the actual equatorial rotational velocity, $P(\theta|v)$ stands for the conditional probability, and H(x) represents the Heaviside step function, which is 1 for x $\geq$ 0 and 0 otherwise.

\begin{figure}[h]
    \centering
    \includegraphics[width=1\linewidth]{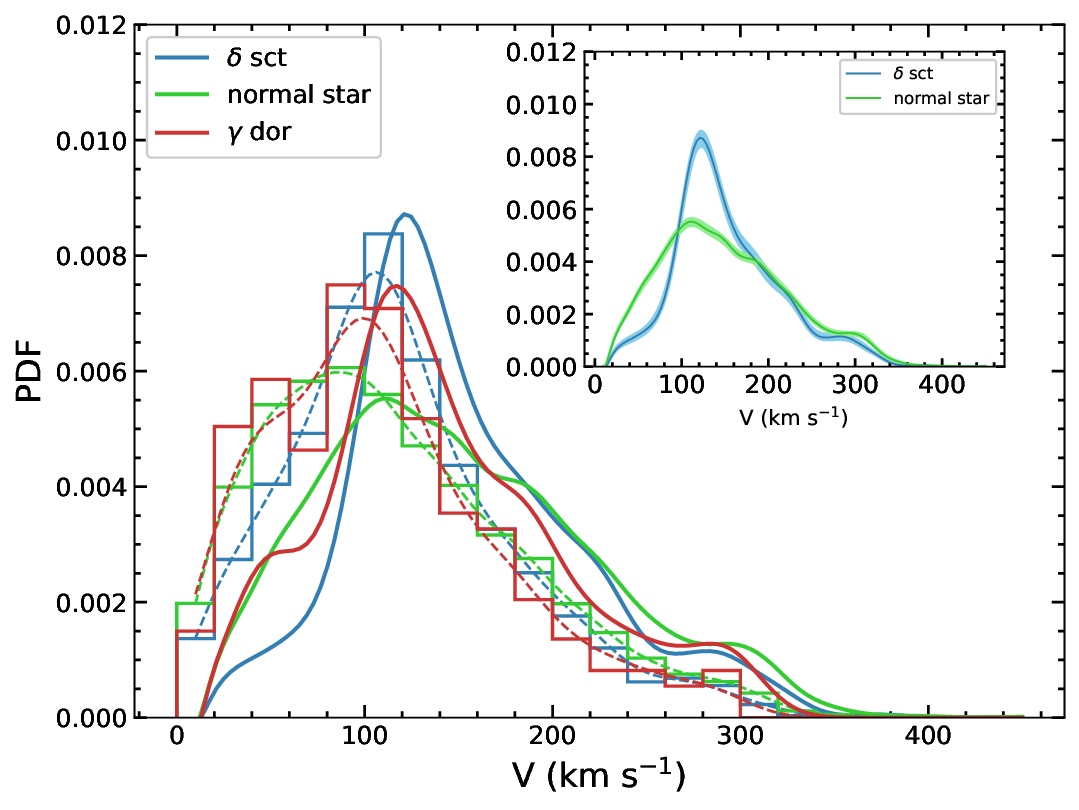}
    \caption{The figure shows the distributions of equatorial rotation velocities for \dsct stars, \gdor stars, and normal stars. Red indicates \gdor stars , blue signifies \dsct stars, and green represents normal stars. The dashed line represents the smoothed distribution of the projected velocity $v\sin i$, while the solid line illustrates the equatorial rotation velocity distribution after correcting for the inclination angle. In the upper right subplot, the shaded region represents the uncertainty of the rotation velocity distribution of \dsct stars and normal stars.}
    \label{fig:3}
\end{figure}

Deconvolution is carried out on the Abel integral through Lucy iteration to derive $\gamma(v)$ \citep{1974AJ.....79..745L,2007A&A...463..671R}. In the study by \citet{2007A&A...463..671R}, a correction was made for the error in $\psi(\theta)$ prior to the correction of the inclination angle, a step that was not included in our investigation. The error in the projected velocity $v\sin i$ is usually less than 10\%. Additionally, when correcting the error in the distribution of rotational velocities in a logarithmic manner, the conversion function from logarithmic distribution to normal distribution is not unique, leading to potential notable discrepancies between the converted $v\sin i$ distribution and the initial distribution. The equatorial rotation velocity distributions for \dsct and \gdor stars are shown in Fig.\hspace{0.25em}\ref{fig:3}.

\begin{figure*}
    \centering
    \includegraphics[width=\linewidth]{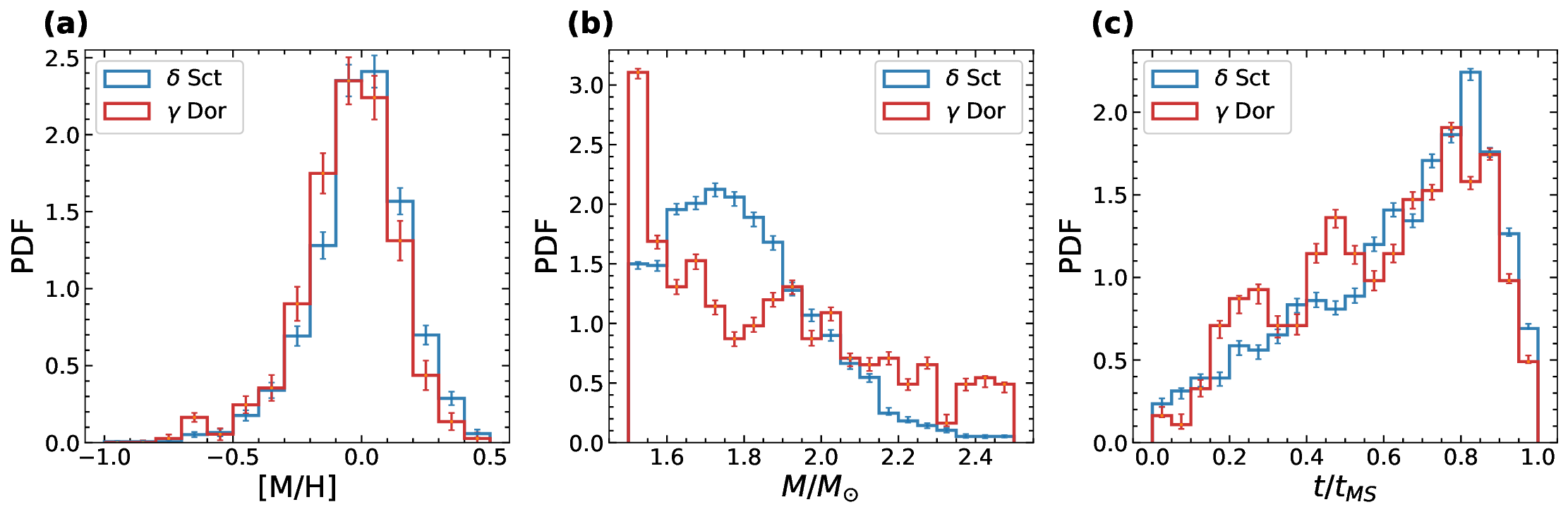}
    \caption{The histogram in the figure illustrates the distribution of metallicity (a), mass (b), and relative ages (c) for \dsct and \gdor stars, with blue representing \dsct stars and red representing \gdor stars. The error bars in the figure represent the uncertainty region at each bin center, derived from bootstrapping.}
    \label{fig:4}
\end{figure*}

The rotational velocity distribution of the normal stars presented in the figure is derived from \citet{2021ApJS..257...22S}. Nevertheless, their final sample still includes a considerable number of pulsating and binary stars, which have not been excluded, making up over thirty percent of the sample for stars with masses below 2.5 \msun. Therefore, we have identified and removed these newly discovered contaminating stars to refine the sample, resulting in a more purer group of normal stars. Using this refined sample, we have determined the equatorial rotational velocity distribution for normal stars as shown in Fig. \ref{fig:3}. In the upper right subplot of Fig. \ref{fig:3}, we display the rotational velocity distributions for \dsct stars and normal stars, along with their 95\% confidence intervals. These intervals were obtained through bootstrapping with 1,000 iterations, sampling from a normal distribution centered on the measured value and using the measurement uncertainty as the standard deviation. The intervals span the 2.5th to 97.5th percentiles of the bootstrap samples, providing a robust estimate by excluding extreme values. The shaded regions indicate these 95\% confidence intervals.

It is clear that the rotational velocity distributions of \dsct stars, \gdor stars, and normal stars are unimodal, aligning with the findings of \citet{2012A&A...537A.120Z,2021ApJ...921..145S} that the rotational velocity distribution of stars with masses less than $2.5$ \msun is unimodal. Furthermore, the rotational velocity distributions of \gdor and \dsct stars exhibit a remarkable similarity, as evidenced by their respective standard deviations of 65 km s$^{-1}$ and 62 km s$^{-1}$, and closely aligned peak velocities at 117 km s$^{-1}$ and 121 km s$^{-1}$. This high degree of consistency suggests that there are virtually no significant differences in their rotational characteristics. The only minor difference is that \gdor stars has a more abundant distribution at the lower velocity end ($v <$ 60 km s$^{-1}$), possibly due to the presence of typically slow-rotating variable stars that have not been entirely excluded from the data. Our findings indicate that the normal stellar rotation distribution includes a 10th percentile velocity of 62 km s$^{-1}$ and a 90th percentile velocity of 267 km s$^{-1}$, with a standard deviation of 73 km s$^{-1}$. On the other hand, the \dsct stars rotation distribution shows a 10th percentile velocity of 87 km s$^{-1}$ and a 90th percentile velocity of 249 km s$^{-1}$. This means that normal stars have a wider spread in rotation velocities compared to \dsct stars, covering more stars with both lower ($v <$ 80 km s$^{-1}$) and higher ($v >$ 200 km s$^{-1}$) velocities, although their peak velocity (110 km s$^{-1}$) is lower than that of \dsct stars. We conducted a Kolmogorov-Smirnov (KS) two-sample test on the rotational velocity distributions between the 10th and 90th percentiles for both \dsct stars and normal stars. The findings reveal that with a sample size of more than 30, the p-value is less than 0.05. When sampling the velocity distribution of \dsct stars and normal stars within the 10th to 90th percentile range using a sample size of 30, the interval for sampling velocities of \dsct stars is 5.4 km s$^{-1}$, whereas for normal stars it is 6.8 km s$^{-1}$. These intervals exceed the velocity uncertainties, which are 3.92 km s$^{-1}$ for \dsct stars and 4.75 km s$^{-1}$ for normal stars. Therefore, there is a statistically significant difference in the distributions between \dsct stars and normal stars.

\section{discussion}\label{sec:discussion}
For a normal single star that is not affected by the interaction between binary stars, the rotation velocity is influenced by the initial angular momentum, mass, and age of the star \citep{2013A&A...553A..24G}. For pulsating stars, the pulsation factor must also be taken into account due to the coupling effect between rotation and pulsation. To thoroughly account for all factors influencing the rotation of \dsct and \gdor stars, we analyzed their mass, age, and metallicity distributions. Fig.\hspace{0.25em}\ref{fig:4} illustrates the distributions of mass, relative age, and metallicity for \dsct and \gdor stars, with blue indicating \dsct stars and red indicating \gdor stars.  The error bars in Fig.\hspace{0.25em}\ref{fig:4} illustrate the uncertainty intervals at the center of each bin, calculated using the same method described above. The masses and ages of the sample stars, similarly sourced from \citep{2021ApJS..257...22S}, were estimated through the evolutionary tracks in the $\log T_\mathrm{eff}-\log L$ plane, utilizing the PARSEC 1.2S stellar evolution code \citep{2012MNRAS.427..127B}, which accounts for the chemical homogenization of gas within convective zones. The chemical composition in the PARSEC 1.2S stellar evolution model (adopting solar metallicity) is based on \citet{2011SoPh..268..255C}. The opacity sources (OPAL) used in the model are as follows: in the high-temperature regime, for $4.2 \leq \log T \leq 8.7$, opacity data is derived from \citet{1996ApJ...464..943I}; in the low-temperature regime, for $3.2 \leq \log T \leq 4.1$, opacity data is sourced from \citet{2009A&A...508.1539M}. In the transition region, a linear interpolation between these two opacity tables is applied to ensure continuity in the model. The electron thermal conductivity is provided by \citet{2008ApJ...677..495I}.

The PARSEC 1.2S model assumes a non-rotating stellar framework, while incorporating both core convective overshooting and envelope convective overshooting effects. For stars with masses between $1.5$ and $2.5 \, M_{\odot}$, core convective overshooting is particularly important, as these stars exceed the threshold mass $M_{\text{o2}}$, above which overshooting reaches its maximum efficiency. In this mass range, the overshooting mixing length is set to $l_{\text{ov}} = 0.25 H_p$, where $H_p$ is the local pressure scale height. During the main sequence, the influence of envelope convective overshooting can be neglected; thus, core convective overshooting alone has a significant impact on the stellar evolutionary track. Core convective overshooting increases the effective core mass, altering the stellar evolutionary path by shifting the star’s position in the Hertzsprung–Russell (HR) diagram toward higher luminosities and temperatures. It should be noted that the core convective overshooting in the PARSEC 1.2S model does not correspond to convective penetration, as it primarily describes an extension of convective mixing beyond the formal convective boundary without deep penetration into the radiative region. However, if core convective overshooting is neglected in the model, systematic biases are introduced when fitting stellar masses and ages, typically leading to overestimated ages and masses.

Additionally, the study by \citet{2022A&A...665A.126N} demonstrates that rotation introduces complexities in the evolution of intermediate- to high-mass stars. At early evolutionary stages, geometric effects due to rotation reduce both luminosity and temperature. However, as stellar evolution progresses, rotational mixing supplies additional hydrogen to the core, gradually increasing the core mass and ultimately enhancing stellar luminosity. Consequently, the impact of rotation on mass and age estimations is not a simple systematic offset; rather, it varies across evolutionary stages, necessitating further detailed modeling. Because our sample stars have a narrow range in $\mathrm{T_{eff}}$ and $\log g$, the choice of physics in evolutionary models will only systematically affect their mass and age, without altering the statistical analysis of relative properties such as the dependence of rotation velocity on mass and age.

\begin{figure*}
    \centering
    \includegraphics[width=\linewidth]{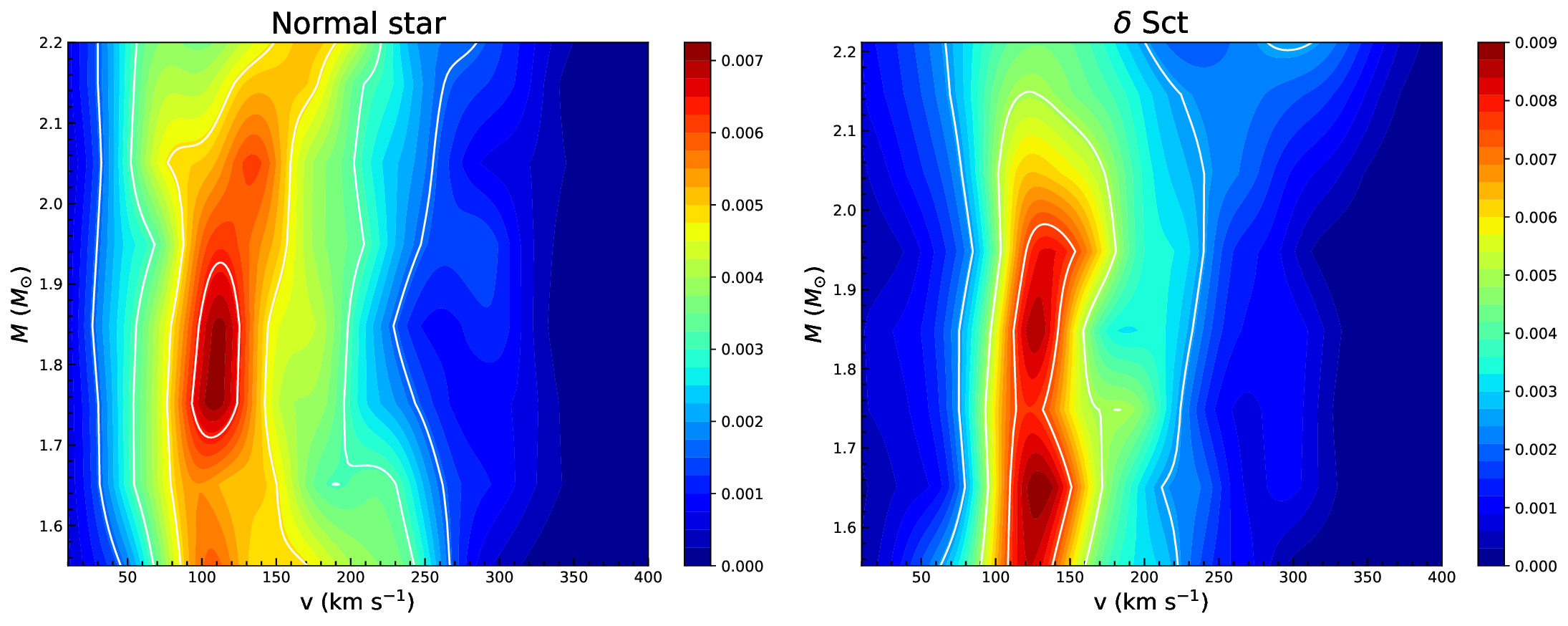}
    \caption{The figure shows the distribution of the equatorial rotation velocities as a function of stellar mass, with the left panel for normal stars after excluding recently confirmed contaminations, and the right panel for pure \dsct stars. The color gradient indicates the density within the normalized one-dimensional distributions, with the isodensity contours depicted as white lines.}
    \label{fig:5}
\end{figure*}

\begin{figure*}
    \centering
    \includegraphics[width=\linewidth]{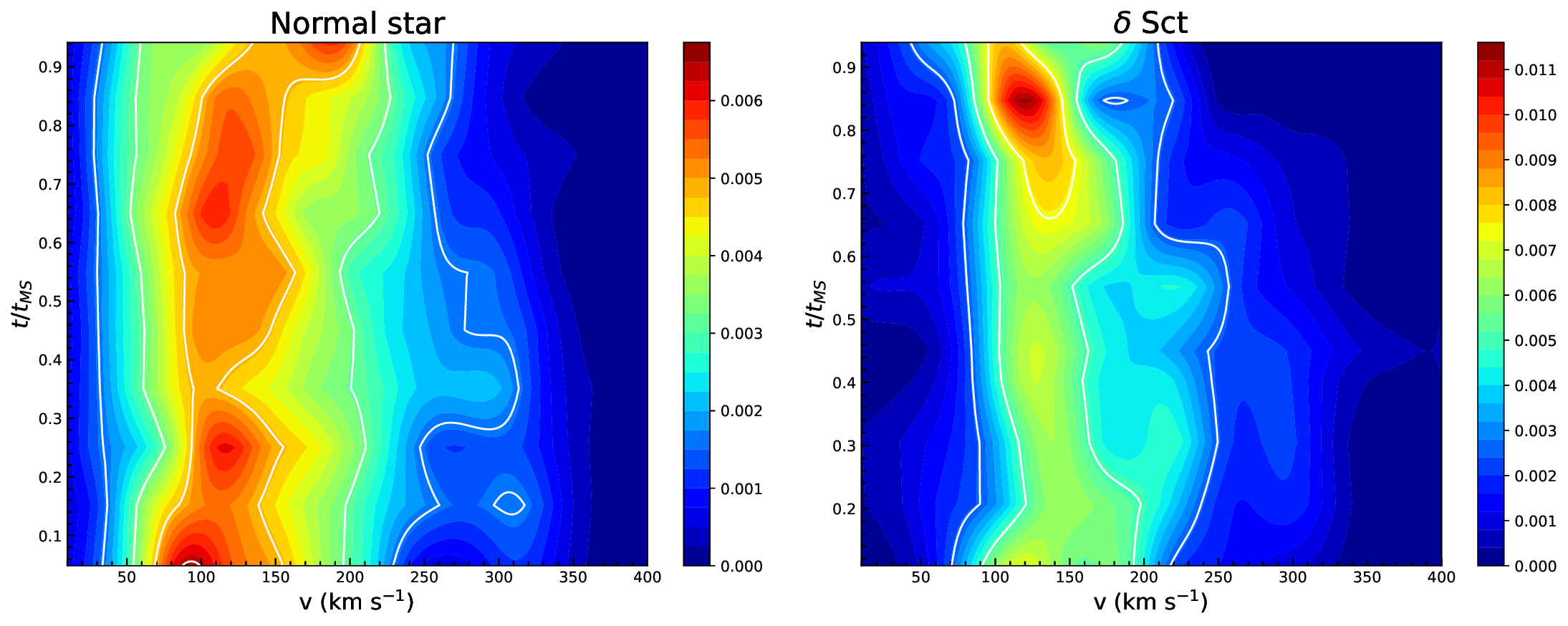}
    \caption{As shown in Fig.\hspace{0.25em}\ref{fig:5}, but it depicts the distribution of equatorial rotation velocities as a function of relative age $t/t_{MS}$ .}
    \label{fig:6}
\end{figure*}

\citet{2021ApJ...921..145S} found a dependence of rotation in early-type stars on metallicity, where in metal-rich ([M/H] $>$ 0.2,  [M/H] $=0$ corresponds to Z=0.0152) environments, a slow rotation is observed at the lower mass ($M < 2.5$ \msun ) end, and a bimodal distribution is evident at the higher mass end. However, metal-poor environments ([M/H] $<-0.2$) do not show a dependence on mass, despite a strong coupling between metallicity and age within this sample. Upon examining the sample, we found that the proportion of metal-rich stars among the normal stars with slow rotation ($v <$ 80 km s$^{-1}$) is nearly double the proportion of metal-rich stars in the overall population of normal stars. In Fig.\hspace{0.25em}\ref{fig:4}a, it is evident that there is no significant difference in the metallicity distributions of the \dsct and \gdor stars, which are mainly concentrated within the range of “normal metal” ($-0.2 < $ [M/H] $ < 0.2$). Consequently, we consider the differential impact of metallicity on the rotation of \dsct and \gdor stars to be relatively negligible.

As shown in Fig.\hspace{0.25em}\ref{fig:4}b, it is evident that while both \dsct and \gdor stars span a broad range of masses across all selected intervals, \dsct stars are predominantly found in the higher mass range compared to \gdor stars. In particular, \dsct stars are mainly concentrated around 1.8 \msun, whereas \gdor stars have the highest likelihood near 1.5 \msun. It is noteworthy that our sample has a lower mass cutoff at 1.5 \msun, whereas the mass of \gdor stars can extend down to 1.3 \msun. Consequently, the peak in the mass distribution of \gdor stars, as presented in the figure, does not necessarily correspond to 1.5 \msun but may instead reside at a lower value. However, the model used to derive stellar parameters for our sample is limited to stars with temperatures between 7000 K and 15,000 K, restricting our ability to extend the sample to lower-mass ranges \citep{2020ApJS..246....9Z,2021ApJS..257...22S}. Utilizing other models would compromise the consistency of the data.
\cite{2021ApJ...921..145S} found that for normal stars with masses below 2.5 \msun, the peak rotational velocity in the velocity distribution increases monotonically with mass. As previously mentioned, their final sample still contains a high proportion of contaminating stars, which may introduce bias into the results. Therefore, by utilizing a purified sample after the removal of contaminants, we have obtained a more reliable relationship function between rotational velocity and mass, as demonstrated in Fig.\hspace{0.25em}\ref{fig:5}.
Normal stars with less than 2.5 \msun show peak velocity increases with mass only above 1.8 \msun, and the higher the mass, the faster the velocity increases, while the lower the mass, the opposite trend occurs. In the right panel of Fig.\hspace{0.25em}\ref{fig:5}, we find that the rotation velocity distribution of pure \dsct  stars is almost independent of mass, which is completely different from that of normal stars, and this difference may be due to the influence of pulsation.

In Fig.\hspace{0.25em}\ref{fig:4}c, it is evident that the age distributions of \dsct and \gdor stars are quite similar, spanning the entire main sequence stage. Notably, while both types are found throughout the main sequence, they predominantly occur in the later stages, with the \gdor stars population tending to be slightly younger than the \dsct stars population. \cite{2012A&A...537A.120Z} found that less massive stars (1.6 \msun $<$ M $<$ 2.4 \msun) undergo an acceleration phase at $t/t_{MS}$ around 0 to 0.3 or 0.4, and maintaining a high-speed state for an extended period afterward with gradual acceleration, which helps explain the relatively low proportion of slow rotation in less massive stars. 

 Similar to how pulsation causes the relationship between rotational velocity distribution and mass in pulsating stars to differ from that in normal stars, the relationship between pulsating stars and evolutionary age might also differ from that of normal stars. We have derived the relationship between the rotational velocity distribution of \dsct stars and normal stars with their relative age $t/t_{MS}$, as shown in Fig.\hspace{0.25em}\ref{fig:6}. It can be distinctly observed that for normal stars, the peak rotational velocity during the late main-sequence phase ($t/t_{MS} >$ 0.7) increases monotonically, whereas for \dsct stars, it decreases monotonically.

 Considering the differences and similarities in the rotational velocity distributions of \dsct stars and normal stars, as well as the varied dependencies of these distributions on mass and age, we have recognized that pulsations indeed exert an influence on the rotational evolution of stars. The peak rotational velocities of \dsct and \gdor stars are marginally higher than those of normal stars, suggesting that pulsations in \dsct and \gdor stars might aid in transporting angular momentum from the interior to the surface. Asteroseismic measurements of near-core rotation frequencies and ages $\gamma$\,Dor stars have shown that angular momentum transport is needed to explain the observed slow-down of the near-core rotation frequency during the main sequence\citep{2019A&A...626A.121O,2021A&A...650A..58M,2023A&A...677A...6M}. However, due to the limitations in sample size, conducting a detailed study on the relationship between the rotation of \gdor stars and their mass and age is not feasible. Nevertheless, given the striking similarity in the rotational velocity distributions between \dsct and \gdor stars, it is plausible to suggest that p-mode pulsations and g-mode pulsations may exert analogous effects on stellar rotation.

On the other hand, despite our efforts to eliminate contaminants from the normal star sample to ensure its purity, it is undeniable that the sample may still contain various stars with distinct rotational properties. This phenomenon is likely one of the significant factors contributing to the broad dispersion observed in the rotational velocity distribution of the normal star sample.

\section{ amplitude-rotation}\label{sec:amp}

\begin{figure*}
    \centering
    \includegraphics[width=\linewidth]{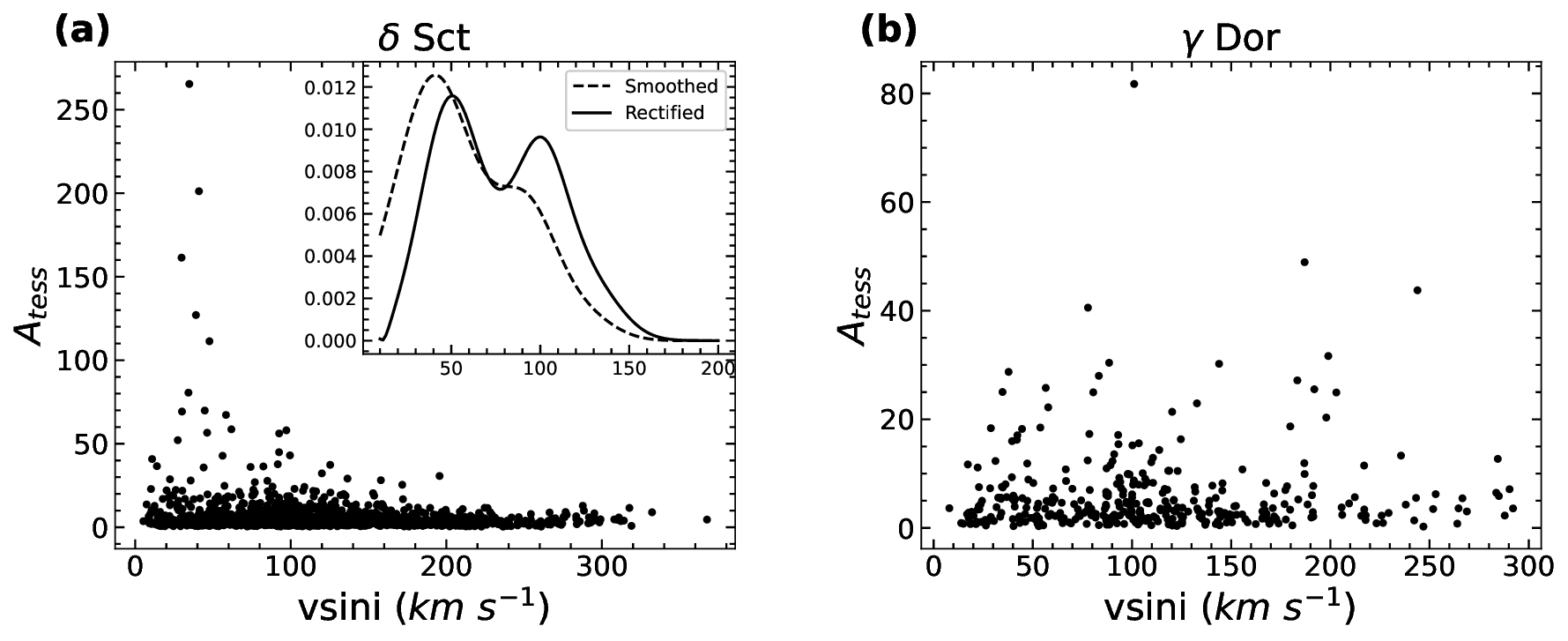}
    \caption{The figure illustrates the relationship between the photometric amplitude (peak to peak) for the primary period in the TESS band and $v\sin i$ for 1534 \dsct (a) and 367 \gdor(b) stars in the sample. And we also present the distribution of rotational velocities with amplitudes exceeding 35 mmag in panel (a), in which we observe a bimodal distribution.}
    \label{fig:7}
\end{figure*}

\begin{figure}
    \centering
    \includegraphics[width=\linewidth]{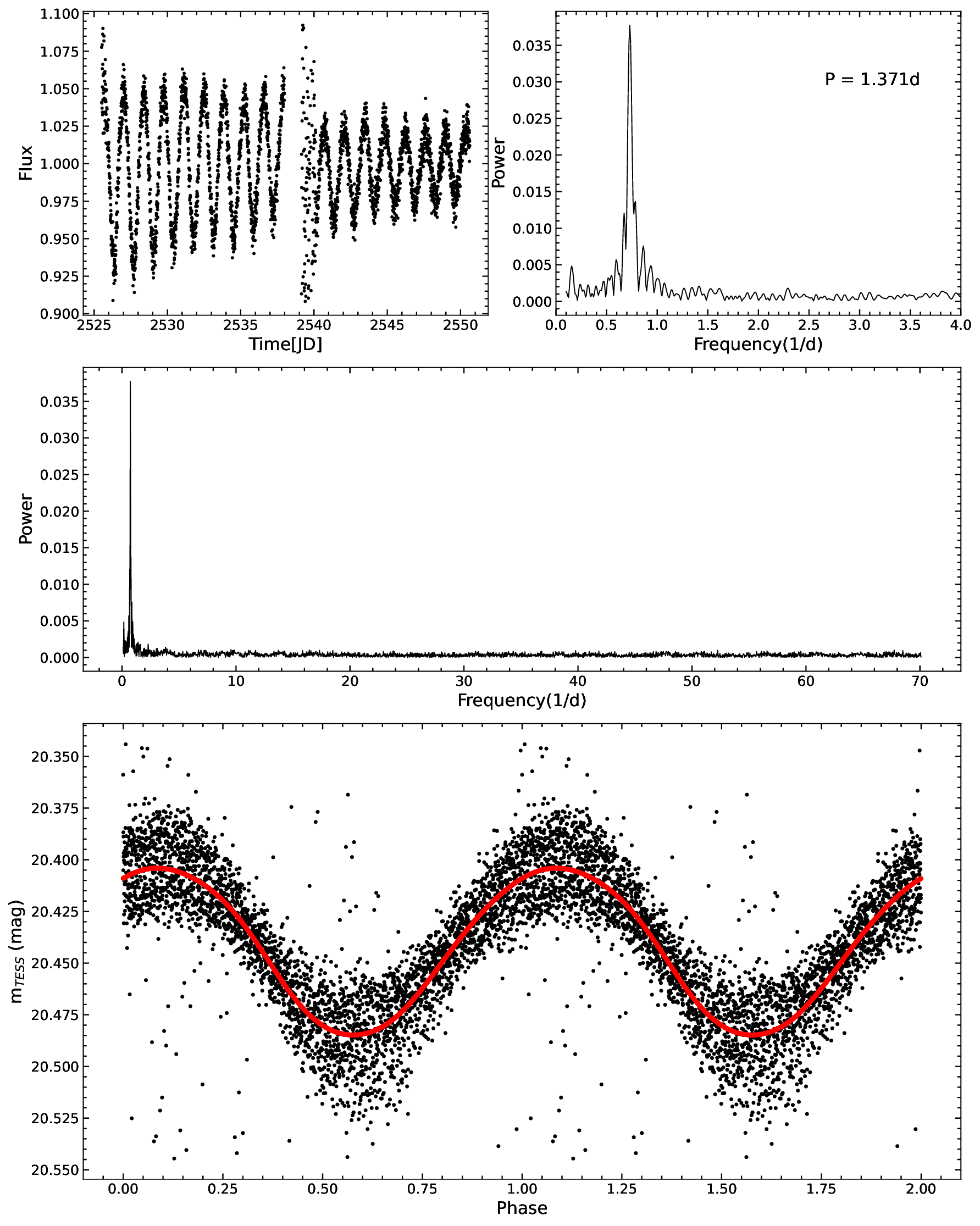}
    \caption{The figure illustrates the light curve, power spectrum, and Fourier fitting outcomes for TIC 83379121.}
    \label{fig:8}
\end{figure}
The earliest-discovered \dsct stars are specifically referred to as high-amplitude \dsct stars (HADS) due to their high-amplitude characteristics, typically satisfying the condition of having an amplitude greater than 0.3 magnitude in the V-band. These unique stars are designated as "dwarf Cepheid variables" due to their exhibited similar Period--Luminosity relation and significant radial pulsation characteristics \citep{2000ASPC..210..373M}. Notably, HADS exhibit a slow rotation, with $v\sin i \leq$ 30km s$^{-1}$, which contrasts sharply with the typical fast rotation exhibited by stars in the same region of the Hertzsprung-Russell diagram. Previously, it was believed that the slow rotation in HADS stars was solely caused by their pulsations being purely radial. However, subsequent studies have revealed the presence of low-amplitude non-radial pulsations in some HADS stars \citep{2003A&A...409.1031P}, although not all of them \citep{2005A&A...440.1097P}. 

Fig. \hspace{0.25em}\ref{fig:7} below demonstrates the correlation between the photometric amplitude of the dominant periods of \dsct and \gdor stars in the TESS band and their $v\sin i$. From the figure, it is evident that high-amplitude (A$_{TESS} > $ 60 mmag ) \dsct stars typically exhibit slower rotational velocities, although they do not necessarily meet the criterion of having $v\sin i$ $\leq$ 30 km s$^{-1}$. However, due to observational constraints, our sample may suffer from significant biases in estimating the rotational velocities of slowly rotating stars \citep{2019ApJ...876..113S}.
The rotational distribution of \dsct stars with amplitudes exceeding 35 mmag (roughly equivalent to 50 mmag in the V-band \citep{2020MNRAS.499.3976P}) reveals a bimodal pattern, indicating the potential presence of two distinct categories of \dsct stars. Those at the lower end of the velocity distribution resemble HADS (dominated by radial pulsations), while those at the higher end belong to the typical low-amplitude \dsct stars (with a large number of non-radial pulsations modes).

Conversely, for low-amplitude \dsct stars, their rotational velocity distribution shows a wide range, encompassing both high and low speeds  \citep{2013AJ....145..132C}. \citet{2007CoAst.150...25B} posits that radial pulsations are strongly influenced by stellar rotation, and only stars with slow rotation can exhibit high-amplitude radial pulsations. In contrast to HADS, low-amplitude \dsct stars with slow rotation may be influenced by non-radial pulsations, thereby unable to generate high-amplitude radial pulsations.

For \gdor stars, we did not find any significant modulation between photometric amplitudes and rotation.
The amplitude of \gdor stars typically does not exceed 0.1 mag (V band), with the high amplitude \gdor stars explained by the superposition of multiple frequencies. Despite the primary frequency amplitude not exceeding 0.1 mag, the total amplitude can reach 0.3 mag \citep{2020MNRAS.499.3976P}.

We examined the star TIC 83379121, which exhibits the largest amplitude within the \gdor class, with its dominant period exceeding 0.1 mag in the V-band. We present its light curve, power spectrum, and Fourier fitting results in Fig. \hspace{0.25em}\ref{fig:8}. Considering its rotational velocity and period, the dominant period is unlikely to be the rotational period. In the figure, we can clearly observe amplitude modulation. Therefore, the high amplitude of this star should be the result of beating formed by the pulsation modes of nearby frequencies\citep{2016MNRAS.460.1970B}.

\section{conclusion}\label{sec:conclusion}
 We have assembled a catalog of variable stars, which includes 1534 \dsct stars, 367 \gdor stars, and 1973 hybrid stars, along with 137 candidate stars. Within the hybrid category, there are 1703 \dshybrid and 270 \gdhybrid stars. Candidate stars are those showing periodic photometric variations but failing to meet our classification standards. Based on their predominant frequency, we have classified 105 as `dsct candidates' and 32 as `gdor candidates'.

After correcting for rotational inclination, we derived the rotational velocity distribution of pure \dsct and \gdor stars, as well as the corrected normal star sample below 2.5 \msun. The rotational velocities of \dsct stars, \gdor stars, and normal stars all exhibit a unimodal distribution. The rotational velocity distributions of \dsct and \gdor stars are strikingly similar, with differences at the low-speed end possibly due to variable stars that have not been completely removed.  The distribution of rotational velocities among normal stars is more dispersed compared to that of pulsating stars, featuring a larger proportion of slowly rotating ($v <$ 80 km s$^{-1}$) and rapidly rotating ($v >$ 200 km s$^{-1}$) stars. Furthermore, the peak velocity of the rotational velocity distribution for normal stars is slower than that of pulsating stars. This finding may suggest that in \dsct and \gdor stars, pulsation facilitates the transfer of angular momentum from the stellar interior to the surface, thereby accelerating the surface rotation. For normal stars, the peak rotational velocity between 1.8 and 2.5 \msun increases monotonically with mass, with a higher mass leading to a faster increase, while \dsct stars does not show a strong mass dependence.  Additionally, normal stars accelerate during the late main-sequence evolution phase ($t/t_{MS} > 0.7$), while \dsct stars decelerates.

The main-sequence stars with masses ranging from 1.5 to 2.5 \msun include chemically peculiar stars (CP stars), high-amplitude and low-amplitude pulsation variables, binary star systems, and cluster member stars. These stars typically exhibit their own distinct rotational characteristics. Consequently, the integrated sample comprising these heterogeneous types of stars exhibits a broader and more dispersed distribution of rotational velocities compared to a sample consisting solely of pulsation variables.

We also found modulation between the amplitude and rotational velocity in \dsct stars, whereas \gdor stars did not exhibit such modulation. High-amplitude \dsct stars always rotate slowly, whereas low-amplitude ones exhibit a broad velocity distribution. This could be attributed to the significant influence of rotation on radial pulsations, where only slowly rotating stars exhibit high-amplitude radial pulsations, while non-radial pulsations can impede the generation of high-amplitude radial pulsations. Unlike \dsct stars, there is no clear relationship between the photometric amplitude and the rotational velocity for \gdor stars. For \gdor stars, the amplitude of a single pulsation mode typically does not exceed 0.1 magnitudes in the V band. When a \gdor star exhibits a large amplitude in a single pulsation mode, it is usually the result of the beating between nearby frequencies, and this situation often leads to noticeable amplitude variations. 

Constrained by the limited sample size and the non-uniform distribution of samples, we are unable to better control variables to exclude factors beyond pulsation modes. We anticipate that additional high-quality samples of main sequence \dsct and \gdor stars in the future will aid in our exploration of the interaction between rotation and pulsation.

\section*{Acknowledgements}
This work was supported by the National Natural Science Foundation of China (NSFC) through grants 12173047, 12322306, 12233009. We also thanked the support from the National Key Research and development Program of China, grants 2022YFF0503404. X. Chen acknowledge support from the Youth Innovation Promotion Association of the Chinese Academy of Sciences (no. 2022055). This paper includes data collected with the TESS mission, obtained from the MAST data archive at the Space Telescope Science Institute (STScI). Funding for the TESS mission is provided by the NASA Explorer Program. STScI is operated by the Association of Universities for Research in Astronomy, Inc., under NASA contract NAS 5–26555. The Guoshoujing Telescope (the Large Sky Area Multi-Object Fiber Spectroscopic Telescope; LAMOST) is a National Major Scientific Project built by the Chinese Academy of Sciences. Funding for the project has been provided by the National Development and Reform Commission. LAMOST is operated and managed by the National Astronomical Observatories, Chinese Academy of Sciences. 

$Software$: PARSEC \citep[1.2S;][]{2012MNRAS.427..127B}, Astropy \citep{2013A&A...558A..33A}, Matplotlib \citep{2007CSE.....9...90H}, TOPCAT \citep{2005ASPC..347...29T}

\bibliography{sample631}
\bibliographystyle{aasjournal}

\end{document}